\documentclass[aps,preprint,floats,epsf,epsfig,nofootinbib,letter]{revtex4}

\usepackage{color}
\usepackage{graphicx}
\usepackage{dcolumn}
\usepackage{bm}

\def\ol{\overline}

\def\be{\begin{eqnarray}}
\def\en{\end{eqnarray}}
\def\non{\nonumber}
\def\CP{{\it CP}~}
\def\vma{{_{V-A}}}

\def\la{\langle}
\def\ra{\rangle}
\def\A{{\cal A}}
\def\B{{\cal B}}

\def\PAP{{P\!A_P}}
\def\PAV{{P\!A_V}}
\def\PEP{{P\!E_P}}
\def\PEV{{P\!E_V}}

\begin{document}

\font\el=cmbx10 scaled \magstep2{\obeylines\hfill January, 2012}
\vskip 1.0 cm

\title{Direct {\it CP} violation in two-body hadronic charmed meson decays}

\author{ Hai-Yang Cheng}
\affiliation{Institute of Physics, Academia Sinica, Taipei, Taiwan 11529, ROC}

\author{ Cheng-Wei Chiang}
\affiliation{Department of Physics and Center for Mathematics and Theoretical Physics,
National Central University, Chungli, Taiwan 32001, ROC}
\affiliation{Institute of Physics, Academia Sinica, Taipei, Taiwan 11529, ROC}
\affiliation{Physics Division, National Center for Theoretical Sciences, Hsinchu, Taiwan 30013, ROC}

\bigskip
\begin{abstract}
\bigskip
Motivated by the recent observation of \CP violation in the charm sector by LHCb, we study direct \CP asymmetries in the standard model (SM) for the singly Cabibbo-suppressed two-body hadronic decays of charmed mesons using the topological-diagram approach. In this approach, the magnitude and the phase of topological weak annihilation amplitudes which arise mainly from final-state rescattering can be extracted from the data. Consequently, direct \CP asymmetry $a_{dir}^{\rm (tree)}$ at tree level can be {\it reliably} estimated. In general, it lies in the range $10^{-4}<a_{dir}^{\rm (tree)}<10^{-3}$. Short-distance QCD penguins and penguin annihilation are calculated using QCD factorization. Their effects are generally small, especially for $D\to VP$ modes. Since weak penguin annihilation receives long-distance contributions from the color-allowed tree amplitude followed by final-state rescattering, it is expected to give the dominant contribution to the direct \CP violation in the decays $D^0\to K^+K^-$ and $D^0\to \pi^+\pi^-$ in which $a_{dir}^{\rm (tree)}$ is absent. The maximal $\Delta a_{CP}^{\rm dir}$, the direct \CP asymmetry difference between the above-mentioned two modes,  allowed in the SM is around $-0.25\%$, more than $2\sigma$ away from the current world average of $-(0.645\pm 0.180)\%$.

\end{abstract}


\pacs{Valid PACS appear here}

\maketitle
\small
%
%
\section{Introduction \label{sec:intro}}

Recently the LHCb Collaboration has reported a result of a nonzero value for the difference between the time-integrated \CP asymmetries of the decays $D^0\to K^+K^-$ and $D^0\to\pi^+\pi^-$ \cite{LHCb}
\be
\Delta A_{CP}\equiv A_{CP}(K^+K^-)-A_{CP}(\pi^+\pi^-)=-(0.82\pm0.21\pm0.11)\% \qquad {\rm (LHCb)}
\en
based on 0.62 fb$^{-1}$ of 2011 data. The significance of the measured deviation from zero is 3.5$\sigma$. However, based on a data sample corresponding to the integrated luminosity of $5.9$ fb$^{-1}$, the CDF Collaboration \cite{CDF} obtained $A_{CP}(\pi^+\pi^-)=(0.22\pm0.24\pm0.11)\%$ and $A_{CP}(K^+K^-)=-(0.24\pm0.22\pm0.09)\%$, and hence
\be
\Delta A_{CP}=-(0.46\pm0.31\pm0.11)\% \qquad {\rm (CDF)} \ .
\en
The time-integrated asymmetry can be written to first order as
\be
A_{CP}(f)=a_{CP}^{\rm dir}(f)+{\la t\ra\over\tau} a_{CP}^{\rm ind} \ ,
\en
where $a_{CP}^{\rm dir}$ is the direct \CP asymmetry, $a_{CP}^{\rm ind}$ is the indirect \CP asymmetry, $\la t\ra$ is the average decay time in the sample and $\tau$ is the $D^0$ lifetime.
A global fit to the data of $\Delta A_{CP}$ is consistent with no \CP violation only at 0.128\% CL \cite{HFAG}. The central values and $\pm1\sigma$ errors for the individual parameters are:
$a_{CP}^{\rm ind} = -(0.019 \pm 0.232 )\%$ and $\Delta a_{CP}^{\rm dir}=
-(0.645\pm 0.180 )\%$.

Whether the first evidence of \CP violation in the charm sector observed by LHCb is consistent with the standard model (SM) or implies new physics will require further analysis of more data and improved theoretical understanding.  For some early and recent theoretical studies, see Refs.~\cite{Quigg:1979ic,Golden:1989qx,Hinchliffe:1995hz,Ligeti,Kagan,Zhu,Rozanov,Nir,PU,Bigi,BigiCDF}.

It is known that in order to induce direct \CP violation, one needs at least two distinct decay amplitudes with non-trivial strong and weak phase differences. In $B$ physics, there exist several QCD-inspired approaches describing the nonleptonic $B$ decays, such as QCD factorization (QCDF) \cite{BBNS99}, pQCD \cite{pQCD} and soft collinear effective theory \cite{SCET}. Even so, one needs to consider $1/m_b$ power corrections in order to explain the observed direct \CP asymmetries in $B$ decays. For example, in QCDF it is necessary to take into account two different types of power correction effects in order to resolve the \CP puzzles and rate deficit problems with penguin-dominated two-body decays of $B$ mesons  and color-suppressed tree-dominated $\pi^0\pi^0$ and $\rho^0\pi^0$ modes: penguin annihilation and soft corrections to the color-suppressed tree amplitude \cite{CCBud}.

The situation is far worse in the charm sector as a theoretical description of the underlying mechanism for exclusive hadronic $D$ decays based on QCD is still not yet available.  This is because the mass of the charmed quark, being of order 1.5 GeV, is not heavy enough to allow for a sensible heavy quark expansion.  Indeed, it does not make too much sense to generalize the QCDF and pQCD approaches to charm decays as the $1/m_c$ power corrections are so large that the heavy quark expansion is beyond control. In short, there is no reliable model which allows us to estimate the phases and magnitudes of the decay amplitudes beyond the color-allowed tree amplitude.

Luckily, we do have a powerful tool which provides a model-independent analysis of the charmed meson decays based on symmetry, namely, the diagrammatic approach. In this approach, the flavor-flow diagrams are classified according to the topologies of weak interactions with all strong interaction effects included.  Based on flavor SU(3) symmetry, this model-independent analysis enables us to extract the topological amplitudes and probe the relative importance of different underlying decay mechanisms. It is complementary to the factorization approaches. Analysis based on the flavor-diagram approach indicates a sizable weak annihilation ($W$-exchange or $W$-annihilation) topological amplitude with a large strong phase relative to the tree amplitude. Since weak annihilation and final-state interactions (FSI's) are both of order $1/m_c$ in the heavy quark limit, this means FSI's could play an essential role in charm decays. Indeed, weak annihilation contributions arise mainly from final-state rescattering, and this explains why an approach based on heavy quark expansion in $1/m_c$ is not suitable for charm decays.

The great merit and the strong point of the topological-diagram approach is that the magnitude and the relative strong phase of each individual topological tree amplitude in charm decays can be extracted from the data. \footnote{This is not the case in $B$ decays where one has to make additional assumptions in order to extract individual topological amplitude cleanly from the data.}  This allows us to calculate \CP asymmetry at tree level in a reliable way, granting us an idea about the size of \CP violation in charmed meson decays.

Since there is no tree-level direct \CP violation in $D^0\to K^+K^-$ and $D^0\to \pi^+\pi^-$ decays, the observation of $\Delta A_{CP}$ by LHCb indicates that, contrary to the conventional wisdom, penguin diagrams in singly Cabibbo-suppressed (SCS) decay channels do play a crucial role for \CP violation even though they may not affect the branching fractions. Indeed, this observation is quite natural in the topological approach since, just as the enhancement in weak annihilation diagrams through FSI's, weak penguin annihilation (more specifically, the QCD-penguin exchange diagram to be introduced in the next section) also receives contributions from the color-allowed tree amplitude followed by final-state rescattering.

The time-integrated \CP asymmetry receives indirect \CP-violating contributions denoted by $a_{CP}^{\rm ind}$ through the $D^0$-$\overline D^0$ mixing.
Such a mixing is governed by the mixing parameters $x\equiv (m_1-m_2)/\Gamma$ and $y\equiv (\Gamma_1-\Gamma_2)/(2\Gamma)$ for the mass eigenstates $D_1$ and $D_2$, where $m_{1,2}$ and $\Gamma_{1,2}$ are respective masses and decay widths and $\Gamma$ is the average decay width. It is known that the short-distance contribution to the mixing parameters is very small \cite{Cheng:1982}, of order $10^{-6}$. On the theoretical side, there are two approaches: the inclusive one relying on the $1/m_c$ expansion (see {\it e.g.}, Ref.~\cite{Lenz} for a recent study), and the exclusive one with all final states summed over. In Ref.~\cite{Falk:y}, only the SU(3) breaking effect in phase space was considered for the estimate of $y$. Consequently, the previous estimate of mixing parameters is subject to larger uncertainties. We believe that a better approach is to concentrate on two-body decays and rely more on data than on theory. This is because the measured two-body decays account for about 75\% of hadronic rates of the $D$ mesons. For $PP$ and $VP$ modes, data with good precision for Cabibbo-favored (CF) and SCS decays are now available. For as-yet unmeasured doubly Cabibbo-suppressed (DCS) modes, their rates can be determined from the diagrammatic approach. We obtained $x=(0.10\pm0.02)\%$ and $y=(0.36\pm0.26)\%$ from the $PP$ and $VP$ final states which account for nearly half of the hadronic width of $D^0$ \cite{CC:mixing} .  It is conceivable that when all hadronic states are summed over, one could have $x\sim (0.2-0.4)\%$ and $y\sim (0.5-0.7)\%$.  At any rate, indirect \CP violation is suppressed by the smallness of the mixing parameters $x$ and $y$ and, moreover, it tends to cancel in the difference between $K^+K^-$ and $\pi^+\pi^-$ final states.

There are also a lot of experimental efforts measuring direct \CP asymmetries in $D$ decays with 3-body and 4-body final states. This will involve a more complicated and time-consuming Dalitz-plot analysis. Nevertheless, the analysis will be very rewarding as it will provide us with much more information about the underlying mechanism for \CP violation. We plan to explore this topic elsewhere.

The purpose of this work is to provide a realistic SM estimate of direct \CP violation in two-body hadronic $D$ decays based on the topological-diagram approach. In Sec.~II we recapitulate the essence of the diagrammatic approach and its application to SCS $D\to PP$ and $D\to VP$ decays. The importance of final-state interactions is emphasized. The topological amplitudes are related to the quantities in the QCDF approach in Sec.~III. Sec.~IV is devoted to the calculation of direct \CP asymmetries by taking into account SU(3) breaking and penguin effects. Sec.~V contains our conclusions.

\section{Diagrammatic approach}

\subsection{Analysis for hadronic charm decays}

\begin{figure}[t]
\vspace*{1ex}
\includegraphics[width=2.8in]{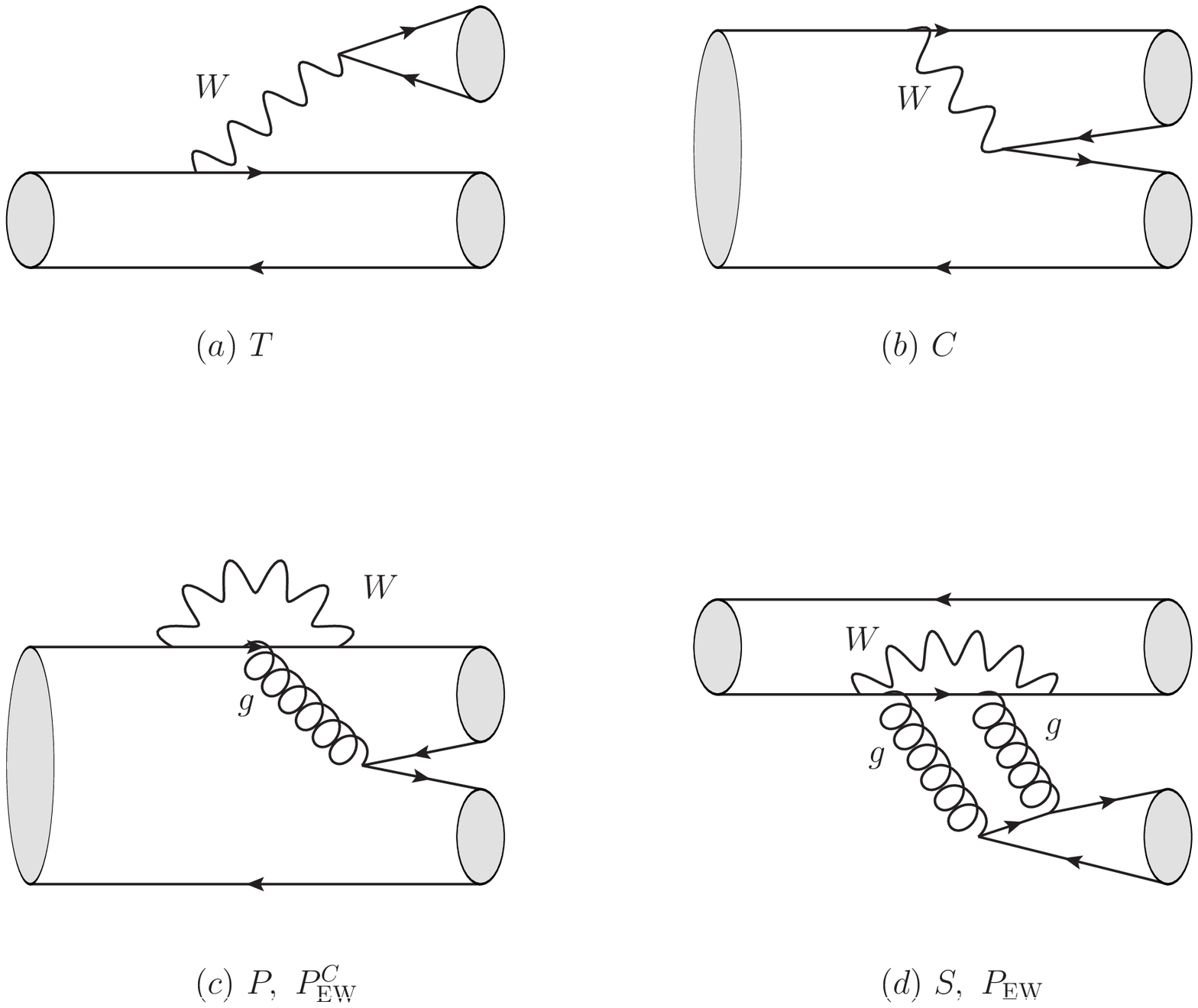}\qquad\includegraphics[width=2.8in]{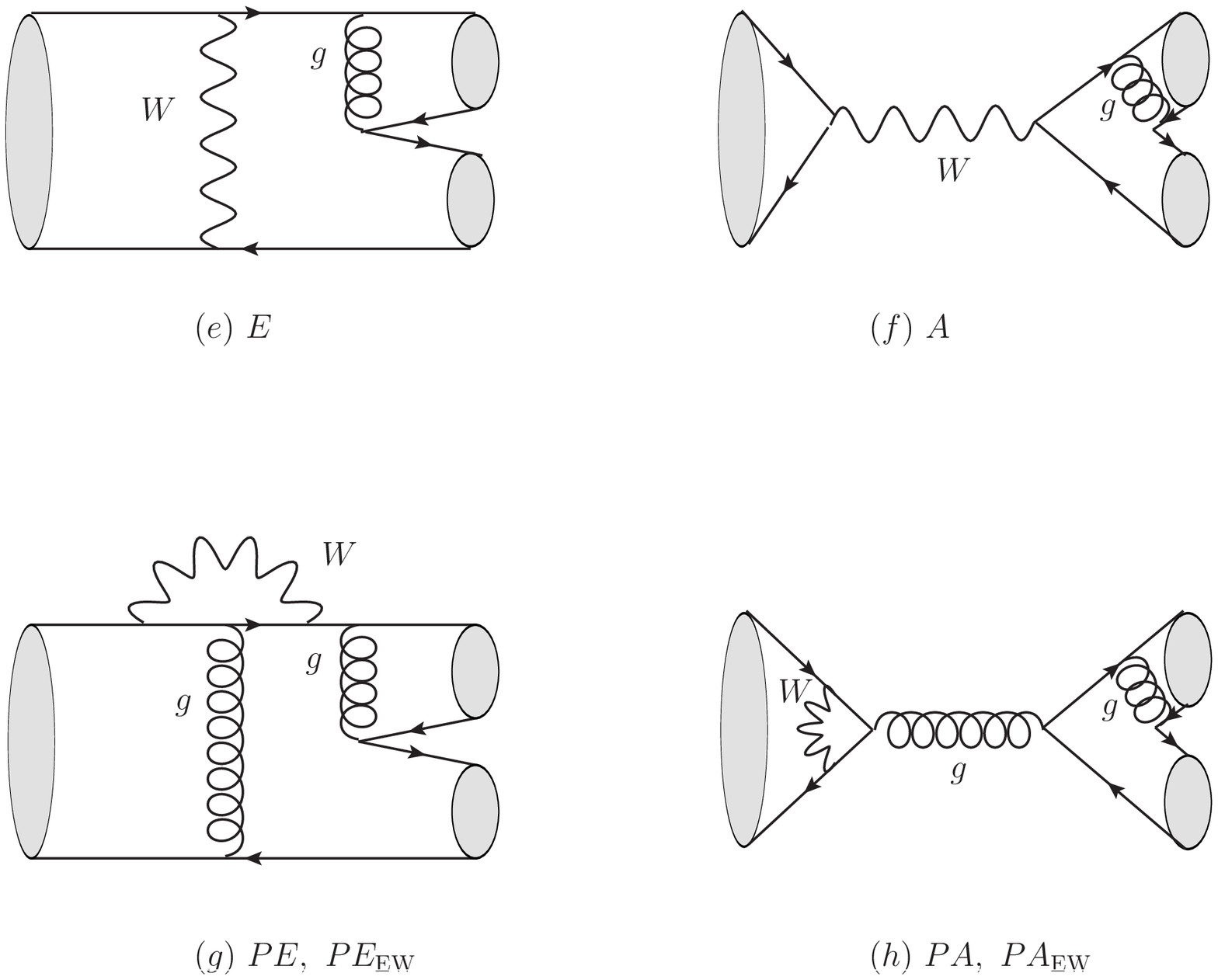}
\vspace*{-1ex}
\caption{ Topology of possible flavor diagrams:
(a) color-allowed tree $T$, (b) color-suppressed tree $C$,
(c) QCD-penguin $P$, (d) singlet QCD-penguin $S$ diagrams with 2 (3) gluon lines for
$M_2$ being a pseudoscalar meson $P$ (a vector meson $V$), where $M_2$ is generally the emitted meson or it contains a quark from the weak vertex for the annihilation diagram, (e) $W$-exchange $E$, (f) $W$-annihilation $A$, (g) QCD-penguin exchange $P\!E$, and
(h) QCD-penguin annihilation $P\!A$ diagrams.
The color-suppressed EW-penguin $P_{\rm EW}^C$ and color-favored EW-penguin
$P_{\rm EW}$ diagrams are obtained by replacing the gluon line from (c) and all the gluon lines from (d), respectively, by a single $Z$-boson or photon line.
The EW-penguin exchange $P\!E_{\rm EW}$ and EW-penguin annihilation $P\!A_{\rm EW}$ diagrams are obtained from (g) and (h), respectively, by replacing the left gluon line by a single $Z$-boson or photon line.
The gluon line of (e) and (f) and the right gluon line of (g) and (h) can be attached to the quark lines in all possible ways.
 } \label{Fig:Quarkdiagrams}
\label{Fig1}
\end{figure}

It has been established sometime ago that a least model-dependent analysis of heavy meson decays can be carried out in the so-called quark-diagram approach \cite{Chau,CC86,CC87}.
In this diagrammatic scenario, the topological diagrams can be classified into three distinct groups as follows (see Fig.~\ref{Fig:Quarkdiagrams}):
\begin{enumerate}
\item tree and penguin amplitudes:
\begin{itemize}
\item $T$, color-allowed external $W$-emission tree amplitude;
\item $C$, color-suppressed internal $W$-emission tree amplitude;
\item $P$, QCD-penguin amplitude;
\item $S$, singlet QCD-penguin amplitude involving SU(3)$_{\rm F}$-singlet mesons (e.g., $\eta^{(\prime)}, ~\omega, ~\phi$);
\item $P_{\rm EW}$, color-favored EW-penguin amplitude; and
\item $P_{\rm EW}^C$, color-suppressed EW-penguin amplitude;
\end{itemize}
\item weak annihilation amplitudes:
\begin{itemize}
\item $E$, $W$-exchange amplitude;
\item $A$, $W$-annihilation amplitude;
($E$ and $A$ are often jointly called ``weak annihilation'' amplitudes.)
\item $P\!E$, QCD-penguin exchange amplitude;
\item $P\!A$, QCD-penguin annihilation amplitude;
\item $P\!E_{\rm EW}$, EW-penguin exchange amplitude; and
\item $P\!A_{\rm EW}$, EW-penguin annihilation amplitude;
($P\!E$ and $P\!A$ are also jointly called ``weak penguin annihilation'' amplitudes) and
\end{itemize}
\item flavor-singlet weak annihilation amplitudes: all involving SU(3)$_{\rm F}$-singlet mesons,
\begin{itemize}
\item $S\!E$, singlet $W$-exchange amplitude;
\item $S\!A$, singlet $W$-annihilation amplitude;
\item $S\!P\!E$, singlet QCD-penguin exchange amplitude;
\item $S\!P\!A$, singlet QCD-penguin annihilation amplitude;
\item $S\!P\!E_{\rm EW}$, singlet EW-penguin exchange amplitude; and
\item $S\!P\!A_{\rm EW}$,  singlet EW-penguin annihilation amplitude.
\end{itemize}
\end{enumerate}
The reader is referred to Ref.~\cite{ChengOh} for details.

It should be stressed that these diagrams are classified purely according to the topologies of weak interactions and flavor flows with all strong interaction effects encoded, and hence they are {\it not} Feynman graphs. All quark graphs used in this approach are topological and meant to include strong interactions to all orders, {\it i.e.}, gluon lines and quark loops are included implicitly in all possible ways.  Therefore, analyses of topological graphs can provide information on FSI's.

The decomposition of the decay amplitudes of SCS $D\to PP$ and $D\to VP$ modes in terms of topological diagrams is displayed in Tables~\ref{tab:CSPP}$-$\ref{tab:CSPV}.  Since we will concentrate exclusively on SCS $D$ decays, the primes on the amplitudes given in Ref.~\cite{ChengChiang} are dropped.  For simplicity, flavor-singlet weak annihilation and weak penguin annihilation amplitudes are neglected in these two tables.

\begin{table}
\caption{Branching fractions and invariant amplitudes for singly Cabibbo-suppressed decays of charmed mesons to two pseudoscalar mesons.  It is understood that the amplitudes with the CKM factor $\lambda_p \equiv V_{cp}^* V_{up}$ are summed over $p = d, s$.  Data are taken from \cite{PDG}.  Predictions based on our best-fitted results in (\ref{eq:PP1}) with exact flavor SU(3) symmetry are given in the last column.
  \label{tab:CSPP}}
\begin{ruledtabular}
\begin{tabular}{l l l c c c c}
 & Mode & Representation
     & ${\cal B}_{\rm exp}$ & ${\cal B}_{\rm theory}$ \\
 & & & ($\times 10^{-3}$)  & ($\times 10^{-3}$) \\
\hline
$D^0$
  & $\pi^+ \pi^-$ & $\lambda_p[(T+E)\delta_{pd} + P^p+P\!E+P\!A]$
     & $1.400 \pm 0.026$ & $2.24 \pm 0.10$ \\
  & $\pi^0 \pi^0$ & ${1\over\sqrt{2}}\lambda_p[(-C+E)\delta_{pd} + P^p+P\!E + P\!A)]$
     & $0.80 \pm 0.05$  & $1.35 \pm 0.05$ \\
  & $\pi^0 \eta $ & $ \lambda_p[-E\delta_{pd}\cos\phi-{1\over\sqrt{2}} C\delta_{ps}\sin\phi + (P^p+P\!E)\cos\phi]$
     & $0.68 \pm 0.07$ & $0.75 \pm 0.02$ \\
  & $\pi^0 \eta' $ & $\lambda_p[-E\delta_{pd}\sin\phi+{1\over\sqrt{2}} C\delta_{ps}\cos\phi + (P^p+P\!E)\sin\phi]$
     & $0.89 \pm 0.14$  & $0.74 \pm 0.02$ \\
  & $\eta\eta $ & $ {1\over\sqrt{2}}\lambda_p\big\{ [(C+E)\delta_{pd}+P^p+P\!E+PA]\cos^2\phi $
     & $1.67 \pm 0.20$ & $1.44 \pm 0.08$ \\
  && \qquad $+
  (-{1\over\sqrt{2}}C\sin 2\phi + 2E\sin^2\phi)\delta_{ps}\big\}$ \\
  & $\eta\eta' $ & $\lambda_p\big\{ {1\over 2}[(C+E)\delta_{pd}+P^p+P\!E+P\!A]\sin 2\phi$
     & $1.05 \pm 0.26$  & $1.19 \pm 0.07$ \\
  && \qquad $+ ({1\over\sqrt{2}}C\cos 2\phi-E\sin 2\phi)\delta_{ps}\big\}$ \\
  & $K^+ K^{-}$ & $\lambda_p[ (T+E)\delta_{ps} + P^p+P\!E+P\!A)]$
     & $3.96 \pm 0.08$  & $1.92 \pm 0.08$ \\
  & $K^0 \ol{K}^{0}$ & $\lambda_p (E'_p+2P\!A)$ \footnotemark[1]
     & $0.692\pm0.116$  & $0$ \\
\hline
$D^+$
  & $\pi^+ \pi^0$ & $\frac{1}{\sqrt{2}}\lambda_d (T+C)$
     & $1.19 \pm 0.06$  & $0.88 \pm 0.10$ \\
  & $\pi^+ \eta $ & $\lambda_p[ \frac{1}{\sqrt{2}}(T+C+2A)\delta_{pd}\cos\phi - C\delta_{ps}\sin\phi$ & $3.53 \pm 0.21$ & $1.48 \pm 0.26$ \\
  & & $\qquad +\sqrt{2}(P^p+P\!E)\cos\phi]$
      \\
  & $\pi^+ \eta' $ & $\lambda_p[ \frac{1}{\sqrt{2}}(T+C+2A)\delta_{pd}\sin\phi +C\delta_{ps}\cos\phi$  & $4.67 \pm 0.29$ & $3.70 \pm 0.37$ \\
  & & $\qquad +\sqrt{2}(P^p+P\!E)\sin\phi]$
     \\
  & $K^+ \ol{K}^{0}$ & $\lambda_p[ A\delta_{pd} + T\delta_{ps} +  P^p+P\!E]$
     & $5.66 \pm 0.32$   & $5.46 \pm 0.53$ \\
\hline
$D_s^+$
  & $\pi^+ K^{0}$ & $\lambda_p[ T\delta_{pd} + A\delta_{ps} +  P^p+P\!E]$
     & $2.42 \pm 0.16$  & $2.73 \pm 0.26$ \\
  & $\pi^0 K^{+}$ & $\frac{1}{\sqrt{2}}\lambda_p[-C\delta_{pd} + A\delta_{ps} + P^p+P\!E]$
     & $0.62\pm 0.21$  & $0.86 \pm 0.09$  \\
  & $K^{+}\eta$ & $\frac{1}{\sqrt{2}}\lambda_p[C\delta_{pd} + A\delta_{ps}
  + P^p+P\!E ]\cos\phi$  & $1.75\pm0.35$  & $0.78 \pm 0.09$ \\
  & & $\qquad -\lambda_p[(T+C+A)\delta_{ps} + P^p+P\!E]\sin\phi$
  \\
  & $K^{+}\eta' $ & $\frac{1}{\sqrt{2}}\lambda_p[C\delta_{pd} + A\delta_{ps}
  + P^p+P\!E]\sin\phi$ & $1.8\pm 0.6$  & $1.07 \pm 0.17$ \\
  & & $\qquad +\lambda_p[(T+C+A)\delta_{ps} + P^p+P\!E] \cos\phi$
   \\
\end{tabular}
\footnotetext[1]{The subscript $p$ in $E'_p$ refers to the quark-antiquark pair popping out of the vacuum in the final state.}
\end{ruledtabular} 
\end{table}

\renewcommand{\arraystretch}{0.85}
\begin{table}
\caption{Same as Table~\ref{tab:CSPP} except for singly Cabibbo-suppressed decays of charmed mesons to one vector and one pseudoscalar mesons.  Due to the lack of information on $A_P$ and $A_V$, no prediction is attempted for $D^+$ and $D_s^+$ decays except $D^+\to\pi^+\phi$.
  \label{tab:CSPV}}
  \scriptsize{
\begin{ruledtabular}
\begin{tabular}{l l l c c c c}
 & Mode & Representation & ${\cal B}_{\rm exp}$
 & ${\cal B}_{\rm theory}$ (A,A1) & ${\cal B}_{\rm theory}$ (S,S1) \\
 & & & $(\times 10^{-3})$ & $(\times 10^{-3})$ & $(\times 10^{-3})$ \\
\hline
$D^0$
  & $\pi^+ \rho^-$ & $\lambda_p [ (T_V+E_P)\delta_{pd}
  + P_V^p+ \PAP + \PEP ]$
     & $4.96 \pm 0.24$ & $3.92 \pm 0.46$ & $5.18 \pm 0.58$ \\
  & $\pi^- \rho^+$ & $\lambda_p [ (T_P+E_V)\delta_{pd}
  + P_P^p+ \PAV + \PEV ]$
     & $9.8 \pm 0.4$ & $8.34 \pm 1.69$ & $8.27 \pm 1.79$ \\
  & $\pi^0 \rho^0$ & $\frac12 \lambda_p [(-C_P-C_V+E_P+E_V)\delta_{pd}$
     & $3.72 \pm 0.22$ & $2.96 \pm 0.98$ & $3.34 \pm 0.33$ \\
  && \qquad $ +P_P^p + P_V^p + \PAP + \PAV + \PEP +\PEV ]$ \\
  & $K^+ K^{*-}$ & $\lambda_p[ (T_V+E_P)\delta_{ps} + P_V^p + \PEP + \PAP]$
     & $1.56 \pm 0.12$ & $1.99 \pm 0.24$ & $1.99 \pm 0.22$ \\
  & $K^- K^{*+}$ & $\lambda_p[ (T_P+E_V)\delta_{ps} + P_P^p + \PEV + \PAV]$
     & $4.38 \pm 0.21$ & $4.25 \pm 0.86$ & $3.18 \pm 0.69$ \\
  & $K^0 \ol{K}^{*0}$ & $\lambda_p( E_V\delta_{pd}+ E_P\delta_{ps} +  \PAP + \PAV)$
     & $<1.5$ & $0.29 \pm 0.22$ & $0.05 \pm 0.06$ \\
  & $\ol{K}^0 K^{*0}$ &  $\lambda_p( E_P\delta_{pd}+ E_V\delta_{ps} +  \PAP + \PAV)$
     & $<0.84$ & $0.29 \pm 0.22$ & $0.05 \pm 0.06$ \\
  & $\pi^0 \omega$ & $\frac{1}{2}\lambda_p [ (-C_V+C_P-E_P-E_V)\delta_{pd}
  + P_P^p + P_V^p + \PEP + \PEV]$
     & $<0.26$ & $0.10 \pm 0.18$ & $1.01 \pm 0.18$ \\
  & $\pi^0 \phi$ & $\frac{1}{\sqrt{2}} \lambda_s C_P$
     & $1.31 \pm 0.17$ & $1.22 \pm 0.08$ & $1.11 \pm 0.05$ \\
  & $\eta \omega$ & $\frac12 \lambda_p[ (C_V+C_P+E_V+E_P)\delta_{pd}\cos\phi
   - C_V\delta_{ps}\sin\phi$
     & $2.21\pm0.23$ \footnotemark[1] & $3.08 \pm 1.42$ & $3.94 \pm 0.61$ \\
  & & $\qquad +(P_P^p + P_V^p +\PEP + \PEV + \PAP + \PAV) \cos\phi]$ &&&& \\
  & $\eta\,' \omega$ & $\frac12 \lambda_p[ (C_V+C_P+E_V+E_P)\delta_{pd}\sin\phi + C_V\delta_{ps}\cos\phi$
     &--- & $0.07 \pm 0.02$ & $0.15 \pm 0.01$ \\
   & & $\qquad +(P_P^p + P_V^p +\PEP + \PEV + \PAP + \PAV) \sin\phi]$ &&&& \\
  & $\eta \phi$ & $\lambda_p \big\{ [{1\over\sqrt{2}}C_P\cos\phi - (E_V+E_P)\sin\phi ]\delta_{ps}
  + (\PAP + \PAV) \sin\phi \big\}$
     & $0.14\pm0.05$ & $0.31 \pm 0.10$ & $0.41 \pm 0.08$ \\
  & $\eta \rho^0$ & ${1\over 2}\lambda_p[ (C_V-C_P-E_V-E_P)\delta_{pd}\cos\phi
  - \sqrt{2}C_V\delta_{ps}\sin\phi$
     & --- &  $1.11 \pm 0.86$ & $1.17 \pm 0.34$ \\
    & & $\qquad + (P_P^p + P_V^p + \PEP + \PEV) \cos\phi]$ &&&& \\
  & $\eta\,' \rho^0$ & ${1\over 2}\lambda_p[ (C_V-C_P-E_V-E_P)\delta_{pd}\sin\phi
  + \sqrt{2}C_V\delta_{ps}\cos\phi$
     & --- & $0.14 \pm 0.02$ & $0.26 \pm 0.02$ \\
      & & $\qquad + (P_P^p + P_V^p + \PEP + \PEV) \sin\phi]$ &&&& \\
\hline
$D^+$
  & $\pi^+ \rho^0$ & $\frac{1}{\sqrt{2}}\lambda_p[ (T_V+C_P-A_P+A_V)\delta_{pd}
  + P_V^p - P_P^p + \PEP - \PEV]$
     & $0.81 \pm 0.15$ &  \\
  & $\pi^0 \rho^+$ & $\frac{1}{\sqrt{2}}\lambda_p[ (T_P+C_V+A_P-A_V)\delta_{pd}
  + P_P^p - P_V^p + \PEV - \PEP]$
     &--- &  \\
  & $\pi^+ \omega$ & $\frac{1}{\sqrt{2}}\lambda_p[ (T_V+C_P+A_P+A_V)\delta_{pd}
  + P_P^p + P_V^p + \PEP + \PEV]$
     &$<0.34$  &  \\
  & $\pi^+ \phi$ & $\lambda_s C_P$
     &$5.42^{+0.16}_{-0.18}$  & $6.21 \pm 0.43$ & $5.68 \pm 0.28$ \\
  & $\eta \rho^+$ & ${1\over\sqrt{2}}\lambda_p [(T_P+C_V+A_V+A_P)\delta_{pd}\cos\phi
   - \sqrt{2}C_V\delta_{ps}\sin\phi$
     & $$ &  \\
  &&$\qquad + (P_P^p + P_V^p + \PEP + \PEV) \cos\phi]$&&&& \\
  & $\eta\,' \rho^+$ & ${1\over\sqrt{2}}\lambda_p[ (T_P+C_V+A_V+A_P)\delta_{pd}\sin\phi
  + \sqrt{2}C_V\delta_{ps}\cos\phi$
     & $$ &  \\
 &&$\qquad + (P_P^p + P_V^p + \PEP + \PEV) \sin\phi]$&&&& \\
  & $K^+ \ol{K}^{*0}$ & $\lambda_p (A_V\delta_{pd} + T_V\delta_{ps} +  P^p_V + \PEP)$
     & $3.68^{+0.14}_{-0.21}$ &  \\
  & $\ol{K}^0 K^{*+}$ & $\lambda_p( A_P\delta_{pd} + T_P\delta_{ps} +  P_P^p + \PEV)$
     & $32 \pm 14$ &  \\
\hline
$D_s^+$
  & $\pi^+ K^{*0}$ & $\lambda_p( T_V\delta_{pd} + A_V\delta_{ps} + P_V^p + \PEP)$
     & $2.25 \pm 0.39$ &  \\
  & $\pi^0 K^{*+}$ & $\frac{1}{\sqrt{2}}\lambda_p\left[ C_V\delta_{pd} - A_V\delta_{ps} - P_V^p - \PEP \right]$
     & --- &  \\
  & $K^+ \rho^0$ & $\frac{1}{\sqrt{2}}\lambda_p\left[ C_P\delta_{pd} - A_P\delta_{ps} - P_P^p - \PEV \right]$
     & $2.7 \pm 0.5$ &  \\
  & $K^0 \rho^+$ & $\lambda_p( T_P\delta_{pd} + A_P\delta_{ps} + P_P^p + \PEV)$
     & --- &  \\
  & $\eta K^{*+}$ & ${1\over\sqrt{2}}\lambda_p\big\{  (C_V\delta_{pd} +  A_V\delta_{ps}
  + P_V^p + \PEP )\cos\phi$
     & --- &  \\
  &&$\qquad - [ (T_P+C_V+A_P)\delta_{ps} + P_P^p + \PEV ]\sin\phi\big\}$&&&& \\
  & $\eta\,' K^{*+}$
  & ${1\over\sqrt{2}}\lambda_p\big\{ (C_V\delta_{pd} + A_V\delta_{ps} + P_V^p + \PEP) \sin\phi$
     & --- &  \\
  &&$\qquad - [ (T_P+C_V+A_P)\delta_{ps} + P_P^p + \PEV ]\cos\phi \big\}$&  &&& \\
  & $K^+ \omega$ & $\frac{1}{\sqrt{2}}\lambda_p\left[ C_P\delta_{pd} + A_P\delta_{ps}
  + P_P^p + \PEV \right]$
     & $<2.4$ &  \\
  & $K^+ \phi$ & $\lambda_p[(T_V+C_P+A_V)\delta_{ps} + P_V^p + \PEP]$
     & $<0.6$  &  \\
\end{tabular}
\footnotetext[1]{Data from \cite{Kass}.}
\end{ruledtabular} }
\end{table}

The topological amplitudes $T,C,E,A$ are extracted from the CF $D\to PP$ decays to be (in units of $10^{-6}$ GeV) \cite{ChengChiang} (see also \cite{RosnerPP08})
\be \label{eq:PP1}
&& T=3.14\pm0.06, \qquad\qquad\qquad\quad
C=(2.61\pm0.08)\,e^{-i(152\pm1)^\circ}, \non \\
&&  E=(1.53^{+0.07}_{-0.08})\,e^{i(122\pm2)^\circ},
\qquad\quad  A=(0.39^{+0.13}_{-0.09})\,e^{i(31^{+20}_{-33})^\circ}
\en
for $\phi=40.4^\circ$ \cite{KLOE}, where  $\phi$ is the $\eta-\eta'$ mixing angle defined in the flavor basis
\be
 \left(\matrix{ \eta \cr \eta'\cr}\right)=\left(\matrix{ \cos\phi & -\sin\phi \cr
 \sin\phi & \cos\phi\cr}\right)\left(\matrix{\eta_q \cr \eta_s
 \cr}\right),
\en
with $\eta_q={1\over\sqrt{2}}(u\bar u+d\bar d)$ and $\eta_s=s\bar s$.

For $D\to VP$ decays, there exist two different types of topological diagrams since the spectator quark of the charmed meson may end up in the pseudoscalar or vector meson. For reduced amplitudes $T$ and $C$ in $D\to VP$ decays, the subscript $P$ ($V$) implies that the pseudoscalar (vector) meson contains the spectator quark of the charmed meson.  For $E$ and $A$ amplitudes with the final state $q_1\bar q_2$, the subscript $P$ ($V$) denotes that the pseudoscalar (vector) meson contains the antiquark $\bar q_2$.

There are two different ways of extracting topological amplitudes: either
 \be \label{eq:VP1}
 \Gamma(D\to VP)={p_c\over 8\pi m_D^2}\sum_{pol.}|\A|^2
 \en
by summing over the the polarization states of the vector meson, or through the relation
 \be \label{eq:VP2}
 \Gamma(D\to VP)={p^3_c\over 8\pi m_D^2}|\tilde \A|^2,
 \en
by taking the polarization vector out of the amplitude, where $\A= (m_V/m_D)\tilde \A\, (\varepsilon\cdot p_D)$.
There exist two solutions, denoted by (A) and (S), for the $T_V$, $C_P$, and $E_P$ amplitudes, depending on whether Eq.~(\ref{eq:VP1}) or Eq.~(\ref{eq:VP2}) is used to extract the invariant amplitudes.  Assuming that $T_P$ and $T_V$ are relatively real, we obtain two best solutions (A1) and (S1) for $T_P$, $C_V$, and $E_V$ \cite{ChengChiang} (see also Ref.~\cite{RosnerVP})
\be \label{eq:VPamp}
&& (A)~~T_V=4.16^{+0.16}_{-0.17},
\quad C_P=(5.14^{+0.30}_{-0.33})\,e^{i(162\pm3)^\circ},
\quad E_P=(3.09\pm0.11)e^{-i(93\pm5)^\circ} ~, \non \\
&& (S)~~T_V = 2.15^{+0.08}_{-0.09},
\quad C_P=(2.68^{+0.14}_{-0.15})\,e^{i(164\pm3)^\circ},
\quad E_P = (1.69\pm0.06)e^{-i(103\pm4)^\circ} ~, \non \\
&& (A1)~~T_P=8.11^{+0.32}_{-0.43},
\quad C_V=(4.15^{+0.34}_{-0.57})\,e^{i(164^{+36}_{-10})^\circ},
\quad E_V=(1.51^{+0.97}_{-0.69})e^{-i(124^{+57}_{-26})^\circ} ~, \non \\
&& (S1)~~T_P = 3.14^{+0.31}_{-0.50},
\quad C_V=(1.33^{+0.36}_{-0.51})\,e^{i(177^{+16}_{-13})^\circ},
\quad E_V = (1.31^{+0.40}_{-0.47})e^{-i(106^{+13}_{-16})^\circ} ~.
\en
Solutions (A) and (A1) in units of $10^{-6}$ are obtained using Eq.~(\ref{eq:VP2}), while solutions (S) and (S1) in units of $10^{-6}(\varepsilon \cdot p_D)$ are extracted using Eq.~(\ref{eq:VP1}). Note that all solutions are extracted from the CF $D\to VP$ decays and that the $A_P$ and $A_V$ amplitudes cannot be completely determined based on currently available data \cite{ChengChiang}.

Under the flavor SU(3) symmetry, one can use the topological amplitudes extracted from the CF modes to predict the rates for the SCS and DCS decays. The branching fractions of SCS $D\to PP$ decays predicted in this way are shown in the last column of Table~\ref{tab:CSPP}. In general, the agreement with experiment is good except for discrepancies in some modes. For example, the predicted rates for $\pi^+\pi^-$ and $\pi^0\pi^0$ are too large, while those for $K^+K^-$, $\pi^+\eta^{(\prime)}$ and $K^+\eta^{(\prime)}$ are too small compared to experiments. The decay $D^0\to K^0\bar K^0$ is prohibited by the SU(3) symmetry, but the measured rate is comparable to that of $D^0\to \pi^0\pi^0$.

Some Cabibbo-suppressed modes exhibit sizable violation of flavor SU(3) symmetry.  We find that part of the SU(3) breaking effects can be accounted for by SU(3) symmetry violation manifested in the color-allowed and color-suppressed tree amplitudes.  However, in other cases such as the ratio $R\equiv\Gamma(D^0\to K^+K^-)/\Gamma(D^0\to\pi^+\pi^-)$, SU(3) breaking in spectator amplitudes is not sufficient to explain the observed value of $R$.  This calls for the consideration of SU(3) violation in the $W$-exchange amplitudes.

\subsection{Final-state rescattering}

From Eq.~(\ref{eq:PP1}) we see that the color-suppressed amplitude $C$ is not only comparable to the tree amplitude $T$ in magnitude but also has a large strong phase relative to $T$. (It is $180^\circ$ in naive factorization.) The $W$-exchange $E$ is sizable with a large phase of order $120^\circ$. Since $W$-exchange is of order $1/m_c$ in the heavy quark limit, this means that $1/m_c$ corrections are very important in charm decays.  Finally, we see that $W$-annihilation is substantially smaller than $W$-exchange and almost perpendicular to $E$.

In naive factorization, the factorizable weak annihilation amplitudes are usually assumed to be negligible as they are helicity suppressed or, equivalently, the form factors are suppressed at large $q^2=m_D^2$. At first glance, it appears that the factorizable weak annihilation amplitudes are too small to be consistent with experiments at all.  However, in the diagrammatic approach here, the topological amplitudes $E$ and $A$ do receive contributions from the tree and color-suppressed amplitudes $T$ and $C$, respectively, via final-state rescattering (see, {\it e.g.}, Fig.~1 of Ref.~\cite{ChengChiang}). Therefore, even if the short-distance weak annihilation vanishes, long-distance weak annihilation can be induced via inelastic FSI's.

For $D\to PP$ decays, it is expected that the long-distance $W$-exchange is dominated by resonant FSI's as shown in Fig.~1(a) of Ref.~\cite{ChengChiang} owing to the fact that an abundant spectrum of resonances is known to exist at energies close to the mass of the charmed meson. It is easy to draw the relevant diagrams, but the difficulty is how to calculate them. In principle, one can evaluate the final-state rescattering contribution at the hadron level (see, {\it e.g.}, Fig.~5 of Ref.~\cite{ChengChiang}), but it involves many theoretical uncertainties. As stressed in the Introduction section, a great merit of the diagrammatic approach is that the magnitude and phase of the topological $W$-exchange and $W$-annihilation amplitudes can be extracted from the data [see Eq.~(\ref{eq:PP1})]. Moreover, the large magnitude and phase of weak annihilation can be quantitatively and qualitatively understood in the following manner. As emphasized in Refs.~\cite{Zen,Weinberg}, most of the properties of
resonances follow from unitarity alone, without regard to the dynamical mechanism that produces the resonances. Consequently, as shown in Refs.~\cite{Zen,Chenga1a2}, the effect of resonance-induced FSI's can be described in a model-independent manner in
terms of the mass and width of the nearby resonances. It is found that the $E$ and $A$ amplitudes are modified by resonant FSI's as (see, {\it e.g.}, Ref.~\cite{Chenga1a2})
\be \label{eq:E,A}
E=e+(e^{2i\delta_r}-1)\left(e+{T\over 3}\right), \qquad
A=a+(a^{2i\delta_{r}}-1)\left(a+{C\over 3}\right),
\en
with
\be
e^{2i\delta_r}=1-i\,{\Gamma\over m_D-m_R+i\Gamma/2},
\en
where the $W$-exchange amplitude $E$ and $W$-annihilation amplitude $A$ before the resonant FSI's are denoted by $e$ and $a$, respectively. Therefore, even if the short-distance weak annihilation is turned off, a long-distance $W$-exchange ($W$-annihilation) contribution can still be induced from the tree amplitude $T$ ($C$) via FSI rescattering in resonance formation. To see the importance of resonant FSI's, consider the scalar resonance $K^*_0(1950)$ with a mass $1945\pm10\pm 20$ MeV and a width $201\pm 34\pm 79$ MeV, contributing to the $W$-exchange in $D\to K\pi,K\eta$. Assuming $e=0$ in Eq.~(\ref{eq:E,A}), we obtain $E=1.68\times 10^{-6}\,{\rm exp}(i143^\circ)\,{\rm GeV}$, which is close to the ``experimental'' value of $E$ given in Eq.~(\ref{eq:PP1}). This suggests that weak annihilation topologies in $D\to PP$ decays are dominated by nearby resonances via final-state rescattering. Contrary to the $PP$ sector, we have shown in Ref.~\cite{ChengChiang} that weak annihilation in $VP$ systems is dominated by final-state rescattering via quark exchange.

\section{Topological amplitudes and QCD factorization}

Although the topological tree amplitudes $T,C,E$ and $A$ for hadronic $D$ decays can be extracted from the data, we still need information on penguin amplitudes (QCD penguin, penguin annihilation, etc.) in order to estimate \CP violation in the SCS decays. To calculate the penguin effect, we start from the short-distance effective Hamiltonian
\be
{\cal H}_{\rm eff}={G_F\over\sqrt{2}}\left[\sum_{p=d,s}\lambda_p(c_1O_1^p+c_2O_2^p+c_{8g}O_{8g})
-\lambda_b\sum_{i=3}^{6}c_iO_i\right] \ ,
\en
where $\lambda_p\equiv V_{cp}^*V_{up}$ for $p=d,s,b$, and
\be
&& O_1^p=(\bar pc)_{_{V-A}}(\bar up)_\vma, \qquad\qquad\quad O_2^p=(\bar p_\alpha c_\beta)_{_{V-A}}(\bar u_\beta p_\alpha)_\vma, \non \\
&& O_{3(5)}=(\bar uc)\vma\sum_q(\bar qq)_{_{V\mp A}}, \qquad~~
O_{4(6)}=(\bar u_\alpha c_\beta)_\vma\sum_q (\bar q_\beta q_\alpha)_{_{V\mp A}},  \non \\
&& O_{8g}=-{g_s\over 8\pi^2}m_c\,\bar u\sigma_{\mu\nu}(1+\gamma_5)G^{\mu\nu}c \ ,
\en
with $O_3$--$O_6$ being the QCD penguin operators and $(\bar q_1q_2)_{_{V\pm A}}\equiv\bar q_1\gamma_\mu(1\pm \gamma_5)q_2$. The electroweak penguin operators are not included in the Hamiltonian as they can be neglected in practice. For the Wilson coefficients, we take $c_1=1.21$, $c_2=-0.41$, $c_3=0.02$, $c_4=-0.04$, $c_5=0.01$, $c_6=-0.05$ and $c_{8g}=-0.06$ from Ref.~\cite{Ligeti} evaluated at the scale $\mu=m_c$.

To evaluate the hadronic matrix elements of the 4-quark operators, it is necessary to specify a suitable framework for the task. We will work with the QCDF approach \cite{BBNS99,BN}. As shown in details in Ref.~\cite{ChengOh}, various topological amplitudes of $D\to M_1M_2$ decays can be expressed in terms of the quantities calculated in the framework of QCDF as follows:
\be \label{eq:P}
 T
   &=& {G_F \over \sqrt{2}}\,a_1 (M_1M_2)
   X^{(D M_1, M_2)} \ ,  \nonumber \\
 C
   &=& {G_F \over \sqrt{2}}\,a_2 (M_1M_2)
   ~X^{(D M_1, M_2)}  \ ,  \nonumber \\
 E
 &=& {G_F \over \sqrt{2}}\,(i f_D f_{M_1} f_{M_2})
   \left[ b_1 \right]_{M_1 M_2}~, \nonumber \\
 A
 &=& {G_F \over \sqrt{2}}\,(i f_D f_{M_1} f_{M_2})
   \left[ b_2 \right]_{M_1 M_2}~, \nonumber \\
 P^p
   &=& {G_F \over \sqrt{2}}\,[a^p_4(M_1M_2)+\eta\, r_\chi^{M_2}a^p_6(M_1M_2)]
   X^{(D M_1, M_2)}  ~,  \nonumber \\
 P\!E
 &=& {G_F \over \sqrt{2}}\,
   (i f_D f_{M_1} f_{M_2})\left[ b_3 \right]_{M_1 M_2 } ~, \nonumber \\
 P\!A
 &=& {G_F \over \sqrt{2}} \,
   (i f_D f_{M_1} f_{M_2})\left[ b_4 \right]_{M_1 M_2 } ~,
\en
where $\eta=1$ for $M_1M_2=PP,PV$ and $\eta=-1$ for $M_1M_2=VP$. In the above equations, the quantities $b_i$ with $i=1,\cdots,4$ expressed in terms of annihilation amplitudes are defined in Ref.~\cite{BN}, and $X$ is a factorizable matrix element given by
\be \label{eq:X}
X^{(D P_1, P_2)} &\equiv& \langle P_2| J^{\mu} |0 \rangle
  \langle P_1| J'_{\mu} |D \rangle
  =i f_{P_2} (m_{D}^2 -m^2_{P_1}) ~F_0^{D P_1} (m_{P_2}^2) ~, \non \\
X^{(DP, V)} &\equiv & \langle V| J^{\mu} |0 \rangle
  \langle P| J'_{\mu} |D \rangle
  =2 f_V \,m_D\, p_c ~F_1^{DP} (m_{V}^2) ~,  \nonumber \\
 X^{(D V,P)} &\equiv & \langle P| J^{\mu} |0 \rangle
  \langle V| J'_{\mu} |D \rangle
  = 2 f_P \,m_D\, p_c ~A_0^{DV} (m_{P}^2) ~,
\en
with $p_c$ being the center-of-mass momentum of either final state particle.  Here we have followed the conventional Bauer-Stech-Wirbel definition for form factors $F_{0,1}^{DP}$ and $A_0^{DV}$~\cite{BSW}. For annihilation amplitudes, we choose the convention that $M_1$ ($M_2$) contains an antiquark (a quark) from the weak vertex. The chiral factors $r_\chi^{M_2}$ in Eq.~(\ref{eq:P}) are given by
\begin{eqnarray}
 r_\chi^P(\mu) = {2m_P^2 \over m_c(\mu)(m_2+m_1)(\mu)},  \qquad\quad
 r_\chi^V(\mu) = \frac{2m_V}{m_c(\mu)} ~\frac{f_V^\perp (\mu)}{f_V} ~,
\end{eqnarray}
with $f_V^\perp (\mu)$ being the scale-dependent transverse decay constant of the vector meson $V$. The flavor operators $a_i^{p}$ in Eq.~(\ref{eq:P}) are basically the Wilson coefficients in conjunction with short-distance nonfactorizable corrections such as vertex corrections and hard spectator interactions. In general, they have the expressions \cite{BN}
 \be \label{eq:ai}
  a_i^{p}(M_1M_2) =
 \left(c_i+{c_{i\pm1}\over N_c}\right)N_i(M_2)
  + {c_{i\pm1}\over N_c}\,{C_F\alpha_s\over
 4\pi}\Big[V_i(M_2)+{4\pi^2\over N_c}H_i(M_1M_2)\Big]+{\cal P}_i^{p}(M_2),
 \en
where $i=1,\cdots,10$, the upper (lower) signs apply when $i$ is odd (even), $c_i$ are the Wilson coefficients, $C_F=(N_c^2-1)/(2N_c)$ with $N_c=3$, $M_2$ is the emitted meson, and $M_1$ shares the same spectator quark as the $D$ meson. The quantities $V_i(M_2)$ account for vertex corrections, $H_i(M_1M_2)$ for hard spectator interactions with a hard gluon exchange between the emitted meson and the spectator quark of the $D$ meson, and ${\cal P}_i(M_2)$ for penguin contractions. The explicit expressions of $V_i$, $H_i$ and ${\cal P}_i$ can be found in Ref.~\cite{BN}. The quantities $N_i(M_2)$ vanish when $i=6,8$ and $M_2=V$, and are equal to unity otherwise.

In general, the decay amplitude is evaluated at the scale $\mu=m_c(m_c)$. However, as stressed in Ref.~\cite{BBNS99}, the hard spectator and annihilation contributions should be evaluated at the hard-collinear scale $\mu_h=\sqrt{\mu\Lambda_h}$ with $\Lambda_h\approx 500$ MeV. This means $\mu_h\approx$ 0.8\,GeV for $D\to M_1M_2$ decays, which is beyond the regime where perturbative QCD is applicable. Therefore, we shall not consider the spectator contributions to $a_i$.

Let us first consider the penguin amplitude
\be
P^p_{\pi\pi}={G_F \over \sqrt{2}} [a^p_4(\pi\pi)+r_\chi^\pi a^p_6(\pi\pi)]X^{(D\pi, \pi)} \ ,
\en
with
\be
a^p_4(\pi\pi) &=& \left(c_4+{c_3\over N_c}\right)+{c_3\over N_c}\,{C_F\alpha_s\over 4\pi}[V^p_4+{4\pi^2\over N_c}H^p_4]+{\cal P}^p_4,  \non \\
a^p_6(\pi\pi) &=& \left(c_6+{c_5\over N_c}\right)+{c_5\over N_c}\,{C_F\alpha_s\over 4\pi}[V^p_6+{4\pi^2\over N_c}H^p_6]+{\cal P}^p_6.
\en
The strong phase of the QCD penguin amplitude arises from vertex corrections and penguin contractions. The order $\alpha_s$ corrections from penguin contraction read \cite{BN}
\be
{\cal P}^p_4&=&{C_F\alpha_s\over 4\pi N_c}\Bigg\{ c_1\left[ {4\over 3}{\rm ln}{m_c\over \mu}+{2\over 3}-G_{M_2}(s_p)\right]+c_3\left[ {8\over 3}{\rm ln}{m_c\over \mu}+{4\over 3}-G_{M_2}(s_u)-G_{M_2}(1)\right]  \non \\
&& \quad +(c_4+c_6)\left[{16\over 3}{\rm ln}{m_c\over \mu}-G_{M_2}(s_u)-G_{M_2}(s_d)-G_{M_2}(s_s)-G_{M_2}(1)\right] \Bigg\} \ ,  \non \\
{\cal P}^p_6&=&{C_F\alpha_s\over 4\pi N_c}\Bigg\{ c_1\left[ {4\over 3}{\rm ln}{m_c\over \mu}+{2\over 3}-\hat G_{M_2}(s_p)\right]+c_3\left[ {8\over 3}{\rm ln}{m_c\over \mu}+{4\over 3}-\hat G_{M_2}(s_u)-\hat G_{M_2}(1)\right]  \non \\
&& \quad +(c_4+c_6)\left[{16\over 3}{\rm ln}{m_c\over \mu}-\hat G_{M_2}(s_u)-\hat G_{M_2}(s_d)-\hat G_{M_2}(s_s)-\hat G_{M_2}(1)\right] \Bigg\} \ ,
\en
where $s_i=m_i^2/m_c^2$,
\be
G_{M_2}(s)=\int_0^1 dx\,G(s,1-x)\Phi_{M_2}(x), \qquad
\hat G_{M_2}(s)=\int_0^1 dx\,G(s,1-x)\Phi_{m_2}(x),
\en
and $G(s,x)=-4\int^1_0 du\,u(1-u){\rm ln}[s-u(1-u)x]$. Here $\Phi_{M_2}$ ($\Phi_{m_2}$) is the twist-2 (-3) light-cone distribution amplitude for the meson $M_2$.

\subsection{Weak points of QCD-inspired approaches}
Although QCD-inspired approaches such as pQCD \cite{Li} and QCDF \cite{Du,Wu,KCYang,Gao,Grossman} have been applied to charm decays, their results cannot be taken seriously. Since the charm quark mass is not heavy enough, the $1/m_c$ power corrections are so large that a sensible heavy quark expansion is not allowed. The large magnitude and phase of weak annihilation ($E$ or $A$) topological amplitude, which is of order $1/m_c$ in heavy quark limit, is an indicative of the importance of power corrections in charm decays. For the parameters $a_1$ and $a_2$ in $D^0\to K^-\pi^+$ decays, we find
\be
a_1(\bar K\pi)=1.19\,e^{i2.6^\circ} ~,\qquad\quad a_2(\bar K\pi)=0.37\,e^{-i155^\circ} ~,
\en
where the strong phases arise from the vertex correction. In Ref.~\cite{ChengChiang}, we have shown that for $D\to \bar K\pi$ decays
\be
|a_1|=1.22\pm0.02 ~, \qquad |a_2|=0.82\pm0.02 ~, \qquad a_2/a_1=(0.67\pm0.02)e^{-i(152\pm1)^\circ} ~,
\en
as obtained from the topological amplitudes $T$ and $C$ given in Eq.~(\ref{eq:PP1}). It is clear that, while the calculated $a_1$ is consistent with ``experiment'', the predicted $a_2$ in QCDF is too small in magnitude and this again calls for large $1/m_c$ corrections.

While there is no trustworthy theoretical framework for describing the hadronic $D$ decays, in this work we shall use the NLO predictions of QCDF to give a crude estimate of the short-distance penguin contributions.

\section{Direct \CP violation}

\subsection{Tree-level \CP violation}
Direct \CP asymmetry in hadronic charm decays defined by
\be
a_{CP}^{\rm dir}(f)={\Gamma(D\to f)-\Gamma(\overline D\to \bar f)\over \Gamma(D\to f)+\Gamma(\overline D\to \bar f)}
\en
can occur even at the tree level \cite{Cheng1984}. Take the $D_s^+\to K^0\pi^+$ mode as an example. Its decay amplitude reads $\lambda_d(T+P^d+P\!E)+\lambda_s(A+P^s+P\!E)$ (see Table~\ref{tab:CSPP}). The interference between the color-allowed tree and $W$-annihilation amplitudes leads to
the tree-level \CP asymmetry
\be \label{eq:acpKpi}
a_{dir}^{({\rm tree})}(D_s^+\to K^0\pi^+)= {2{\rm Im}(\lambda_d\lambda_s^*)\over |\lambda_d|^2}\,
{{\rm Im}(T^*A)\over |T-A|^2}\approx 1.2\times 10^{-3} \left| {A\over T}\right|\sin\delta_{AT} \ ,
\en
where $\delta_{AT}$ is the strong phase of $A$ relative to $T$ and we have taken into account the fact that the magnitude of $A$ is much smaller than $T$. It is obvious that direct \CP violation in charm decays is CKM suppressed by a factor of $10^{-3}$. Assuming the same topological amplitudes in SCS and CF decays, we then find from Eqs.~(\ref{eq:PP1}) and (\ref{eq:acpKpi}) that $a_{dir}^{({\rm tree})}\approx 10^{-4}$ for $D_s^+\to K^0\pi^+$. Larger direct \CP asymmetry can be achieved in those decay modes with interference between $T$ and $C$ or $C$ and $E$. For example, $a_{dir}^{({\rm tree})}$ is of order $-1.1\times 10^{-3}$ for $D^0\to \omega\eta'$ and of order $(0.6\sim0.7)\times 10^{-3}$ for $D^0\to \pi^0\eta$ and $D^0\to K^0\overline K^{*0}$.

\subsection{Penguin-induced \CP violation}

Direct \CP violation does not occur at the tree level in some of the SCS decays, such as $D^0\to K^+K^-$ and $D^0\to\pi^+\pi^-$. In these two decays, the \CP asymmetry can only arise from the interference between tree and penguin amplitudes
\be \label{eq:pipiacp}
a_{dir}^{({\rm loop})}(\pi^+\pi^-) &=& {2{\rm Im}(\lambda_d\lambda_s^*)\over |\lambda_d|^2}\,
{{\rm Im}[(T^*+E^*+P^{d*}+P\!E+P\!A)(P^s+P\!E+P\!A)]_{\pi\pi}\over |T_{\pi\pi}+E_{\pi\pi}|^2 }   \non \\
&\approx& 1.2\times 10^{-3} \left| {P^s+P\!E+P\!A\over T+E}\right|_{\pi\pi}\sin\delta_{\pi\pi} \ ,
\en
where $\delta_{\pi\pi}$ is the strong phase of $P_{\pi\pi}^s+P\!E_{\pi\pi}+P\!A_{\pi\pi}$ relative to $T_{\pi\pi}+E_{\pi\pi}$. Likewise,
\be  \label{eq:KKacp}
a_{dir}^{({\rm loop})}(K^+K^-)
&\approx& -1.2\times 10^{-3} \left| {P^d+P\!E+P\!A \over T+E}\right|_{_{K\!K}}\sin\delta_{K\!K} \ .
\en

To estimate their \CP asymmetries, we first discuss the rates of $D^0\to \pi^+\pi^-$ and $D^0\to K^+K^-$.  Experimentally, the ratio $R\equiv \Gamma(D^0\to K^+K^-)/\Gamma(D^0\to\pi^+\pi^-)$ is about 2.8 \cite{PDG}, while it should be unity in the SU(3) limit. This is a long-standing puzzle since SU(3) symmetry is expected to be broken at the level of 30\% only. Without the inclusion of SU(3) breaking effects in the topological amplitudes, we see from Table~\ref{tab:CSPP} that the predicted rate of $K^+K^-$ is even smaller than that of $\pi^+\pi^-$ due to less phase space available to the former.  In the factorization approach, SU(3)-breaking effects in the tree amplitudes $T$ are given by
\be
{T_{_{K\!K}}\over T}={f_K\over f_\pi}\,{F_0^{DK}(m_K^2)\over F_0^{DK}(m_\pi^2)} \ , \qquad\qquad  {T_{\pi\pi}\over T}={m_D^2-m_\pi^2\over m_D^2-m_K^2}\,{F_0^{D\pi}(m_\pi^2)\over F_0^{DK}(m_\pi^2)} \ .
\en
Using the form factor $q^2$ dependence
and input parameters given in Ref.~\cite{ChengChiang}, we obtain
\be
 T_{_{K\!K}}/T=1.275 \ , \qquad \qquad T_{\pi\pi}/T=0.96 \ .
\en
This leads to $\B(D^0\to K^+K^-) = (3.4 \pm 0.1) \times 10^{-3}$ and $\B(D^0\to\pi^+\pi^-)= (2.1 \pm 0.1) \times 10^{-3}$ assuming no SU(3) symmetry breaking in $E$, {\it i.e.}, $E_{_{K\!K}}=E_{\pi\pi}=E$. Therefore, SU(3) breaking in spectator amplitudes leads to $R=1.6$ and
is still not sufficient to explain the observed value of $R\approx 2.8$.  This calls for the consideration of flavor symmetry SU(3) violation in the $W$-exchange amplitudes. We have argued in Ref.~\cite{ChengChiang} that the long-distance resonant contribution through the nearby state $f_0(1710)$ could account for SU(3)-breaking effects in the $W$-exchange topology. This has to do with the dominance of the scalar glueball content of $f_0(1710)$ and the chiral suppression effect in the ratio $\Gamma(f_0(1710)\to \pi\bar\pi)/\Gamma(f_0(1710)\to K\bar K)$.  To fit the measured branching fractions we find
\be
E_{_{K\!K}}=1.6\times 10^{-6}\,e^{i108^\circ}{\rm GeV}=1.05\, e^{-i14^\circ}E, \qquad
E_{\pi\pi}=1.3\times 10^{-6}\,e^{i145^\circ}{\rm GeV}=0.85\, e^{i23^\circ}E.
\en

Using the input parameters for the light-cone distribution amplitudes of light mesons, quark masses and decay constants from Refs.~\cite{ChengBud,Bazavov} and form factors from Refs.~\cite{ChengChiang,YLWu},\footnote{More specifically,  we use $f_D=219$ MeV and $f_{D_s}=260$ MeV \cite{Bazavov} for the charmed meson decay constants and those in \cite{YLWu} for $D\to V$ transition form factors.}
we find
\be
\left({P^s\over T+E}\right)_{\pi\pi}=0.35\, e^{-i175^\circ}, \qquad \left({P^d\over T+E}\right)_{K\!K}=0.24\, e^{-i176^\circ}.
\en
Hence, $\delta_{\pi\pi}\approx \delta_{K\!K}=-176^\circ$. From Eqs.~(\ref{eq:pipiacp}) and (\ref{eq:KKacp}), we derive
$a_{dir}^{\rm t+p}(\pi^+\pi^-)=-3.9\times 10^{-5}$ and $a_{dir}^{\rm t+p}(K^+K^-)=2.0\times 10^{-5}$, where $a_{dir}^{\rm t+p}$ denotes the \CP asymmetry arising from the interference between tree and QCD-penguin amplitudes.  QCD-penguin induced \CP asymmetries in $D^0\to \pi^+\pi^-,~ K^+K^-$ are small partially due to the almost trivial strong phases of $\delta_{\pi\pi}$ and $\delta_{K\!K}$.


\begin{table}
\caption{Predictions of direct \CP asymmetries in units of $10^{-3}$, where $a_{dir}^{({\rm tree})}$ denotes \CP asymmetry arising from tree amplitudes. The superscript ${\rm (t+p)}$ denotes tree plus QCD penguin amplitudes, ${\rm (t+pa)}$ for tree plus weak penguin annihilation ($P\!E$ and $P\!A$) amplitudes and ``tot'' for the total amplitude.
For the {\it VP} modes, we use the solutions (A) and (A1) given in Eq.~(\ref{eq:VPamp}) for tree and weak annihilation amplitudes. For weak penguin annihilation, we assume that it is similar to the topological $E$ amplitude [see Eq.~(\ref{eq:PE})].  World averages of experimental measurements are taken from Ref.~\cite{HFAG}.  Due to the lack of information on the topological amplitudes $A_P$ and $A_V$, no prediction is attempted for $D^+\to VP$ and $D_s^+\to VP$ decays.
  \label{tab:CPV}}
  \footnotesize{
\begin{ruledtabular}
\begin{tabular}{l r r r r c| l r r r r }
Decay Mode & $a_{dir}^{({\rm tree})}$
     & $a_{dir}^{({\rm t+p})}$ & $a_{dir}^{({\rm t+pa})}$ & $a_{dir}^{({\rm tot})}$ & Expt($10^{-3}$) & $\!\!\!$ Decay Mode & $a_{dir}^{({\rm tree})}$ & $a_{dir}^{({\rm t+p})}$
     & $a_{dir}^{({\rm t+pa})}$ & $a_{dir}^{({\rm tot})}$ \\
 \hline
$D^0\to \pi^+ \pi^-$ & $0$
     & $-0.04$ & 0.90 & $0.86$ & $2.0\pm2.2$ & $D^0\to \pi^+\rho^-$ & $0$ & $0.10$ & $-0.60$ & $-0.51$ \\
$D^0\to\pi^0 \pi^0$  & $0$
     & $0.25$ & 0.59 & 0.85 & $1\pm48$ & $D^0\to \pi^-\rho^+$ & $0$ & $-0.05$ & $-0.22$ & $-0.27$ \\
$D^0\to \pi^0 \eta $  & $0.63$
     & $0.43$ & 0.03 & $-0.16$ & & $D^0\to \pi^0\rho^0$ & 0 & 0 & $-0.74$ & $-0.74$ \\
$D^0\to \pi^0 \eta' $  & $-0.51$
     & $-0.60$ & 0.09 & $-0.01$ & & $D^0\to K^+K^{*-}$ & 0 & $-0.10$ & 0.60 & $0.50$ \\
$D^0\to \eta\eta $ & $-0.37$
     & $-0.38$ & $-0.70$ & $-0.71$ & & $D^0\to K^-K^{*+}$ & 0 & 0.06 & $0.22$ & $0.29$ \\
$D^0\to \eta\eta' $  & $0.39$
     & $0.44$ & $0.21$ & $0.25$ & & $D^0\to K^0\overline K^{*0}$ & $0.73$& 0.73 & 0.73 & $0.73$ \\
$D^0\to K^+ K^{-}$   & $0$
     & $0.02$ & $-0.50$ & $-0.48$ &  $-2.3\pm1.7$ & $D^0\to \overline K^0 K^{*0}$ & $-0.73$ & $-0.73$ & $-0.73$ & $-0.73$ \\
$D^+\to \pi^+ \pi^0$   & $0$
     & $0$ & 0 & 0 & & $D^0\to \pi^0\omega$ & 0 &  $-0.15$ &  $0.53$ & $0.37$ \\
$D^+\to \pi^+ \eta $  & $0.37$
     & $0.28$ & $-0.56$ & $-0.65$ & $17.4\pm11.5$ \footnotemark[1] & $D^0\to \pi^0\phi$ & $0$ & 0 & 0 & 0 \\
$D^+\to\pi^+ \eta' $  & $-0.21$
     & $-0.26$ & 0.42 & 0.41 & $-1.2\pm11.3$ \footnotemark[1] & $D^0\to \eta\omega$ & 0.19 & 0.19 & $0.50$ & $0.50$ \\
$D^+\to K^+ \ol{K}^{0}$  & $-0.07$
     & $0.08$ & $-0.53$ & $-0.38$ & $-0.9\pm6.3$ & $D^0\to \eta'\omega$ & $-1.07$ & $-1.05$ &  $-0.91$ & $-0.89$ \\
$D_s^+\to \pi^+ K^{0}$  & $0.09$
     & $-0.07$ & 0.69 & 0.52 & $65.3\pm24.6$ & $D^0\to \eta\phi$ & 0 & 0 & 0 & 0 \\
$D_s^+\to \pi^0 K^{+}$  & $0.01$
     & $0.02$ & $0.87$ & 0.88 & $20\pm290$ & $D^0\to \eta\rho^0$ & $-0.53$ & $-0.54$ & $-0.22$ & $-0.23$  \\
$D_s^+\to K^{+}\eta$   & $-0.61$
     & $-0.27$ & $-0.53$ & $-0.19$ & $-200\pm180$ &  $D^0\to \eta'\rho^0$ & $0.59$ & $0.58$ & $0.21$ & $0.20$  \\
$D_s^+\to K^{+}\eta' $  & $0.35$
     & $0.58$ & $-0.63$ & $-0.41$ & $-170\pm370$ &  \\
\end{tabular}
\footnotetext[1]{Data from \cite{Belle}.}
\end{ruledtabular} }
\end{table}
%


\begin{figure}[t]
\begin{center}
\includegraphics[width=0.40\textwidth]{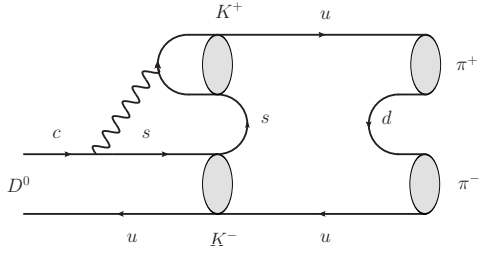}
\vspace{0.0cm}
\caption{Long-distance contribution to $D^0\to \pi^+\pi^-$ through a resonant-like final-state rescattering. It has the same topology as the QCD-penguin exchange topological diagram $P\!E$.
} \label{fig:FSI}
\end{center}
\end{figure}

Short-distance weak penguin annihilation contributions ($P\!E$ and $P\!A$) to charm decays are very small; typically, $P\!E/T$ and $P\!A/T$ are of order $10^{-5}$ and hence are negligible.
Nevertheless, long-distance contributions to SCS decays, for example, $D^0\to \pi^+\pi^-$, can proceed through the weak decay $D^0\to K^+K^-$ followed by a resonant-like final-state rescattering as depicted in Fig.~\ref{fig:FSI}. It has the same topology as the QCD-penguin exchange topological graph $P\!E$.  Just as the weak annihilation topologies $E$ and $A$, it is expected that weak penguin annihilation will receive sizable long-distance contributions from final-state interactions as well. Recall that soft corrections due to penguin annihilation have been proposed to resolve some problems in hadronic $B$ decays, for example, the rate deficit problem for penguin-dominated decays and the \CP puzzle for $\bar B^0\to K^-\pi^+$ \cite{BN}. Hence, we shall assume that $P\!E$, $P\!E_P$ and $P\!E_V$ are of the same order of magnitude as $E$, $E_P$ and $E_V$, respectively. For concreteness, we take (in units of $10^{-6}$)
\be \label{eq:PE}
P\!E=1.6\,e^{i 115^\circ}{\rm GeV}, \qquad P\!E_P=3.1\,e^{-i90^\circ}, \qquad P\!E_V=1.5\,e^{-i120^\circ}.
\en

\subsection{Numerical results and discussion}

The calculated direct \CP asymmetries for various SCS $D\to PP$ and $D\to VP$ decays are summarized in Table~\ref{tab:CPV}.
\footnote{\CP-violating asymmetries in SCS decays of charmed mesons have also been investigated in Ref.~\cite{Buccella}. In this work, final-state interaction effects are studied by assuming that FSI's are dominated by nearby resonances, similar to the work of Refs.~\cite{Zen,Chenga1a2}. The major uncertainties there arise from the unknown masses and widths of the resonances at energies near the charmed meson mass. Noticeable differences between Ref.~\cite{Buccella} and our results for direct \CP asymmetries are that $\Delta a_{CP}^{\rm dir}$ of $D^0\to K^{*+}K^-$ and $D^0\to K^{*-}K^+$ are predicted to have the same sign in Ref.~\cite{Buccella} and likewise for $D^0\to K^+K^-$ and $D^0\to \pi^+\pi^-$, whereas they are of opposite signs in our work.}  Before embarking on \CP violation, we have taken into account the SU(3) breaking effects in tree amplitudes $T$ and $C$ using the factorization approach  so that the predicted branching fractions are consistent with the data shown in Tables~\ref{tab:CSPP} and \ref{tab:CSPV}. Besides the aforementioned example for $D^0\to \pi^+\pi^-$ and $D^0\to K^+K^-$, we found, for example, $T_{\pi^+\eta^{(')}}/T=0.85$, $C_{\pi^+\eta^{(')}}/C=0.93$ for the decays $D^+\to \pi^+\eta^{(')}$ \cite{ChengChiang}. The predicted branching fractions $\B(D^+\to\pi^+\eta)=(3.02 \pm 0.19)\times 10^{-3}$ and $\B(D^+\to\pi^+\eta')=(4.69 \pm 0.21)\times 10^{-3}$ with $A_{\pi^+\eta^{(')}}/A=0.7$ are in agreement with the measured ones $\B(D^+\to\pi^+\eta)=(3.53 \pm 0.21)\times 10^{-3}$ and $\B(D^+\to\pi^+\eta')=(4.67 \pm 0.29)\times 10^{-3}$ (see Table \ref{tab:CSPP}).

In general, tree-level \CP violation lies in the range $10^{-4}<a_{dir}^{\rm (tree)}<10^{-3}$. The largest one occurs in the decay $D^0\to \omega\eta'$ where $a_{dir}^{({\rm tree})}=-1.1\times 10^{-3}$. We would like to accentuate once again that the estimation of tree-level \CP asymmetries in the diagrammatic approach is reliable and trustworthy since the magnitude and phase of the tree topological amplitudes are extracted from the measured data. By an inspection of $a_{dir}^{({\rm tree})}$, one can get an idea about the size of direct \CP violation within the SM.

The QCD penguin effects on \CP asymmetry can be inferred from a comparison between $a_{dir}^{({\rm t+p})}$ and $a_{dir}^{({\rm tree})}$. We see from Table~\ref{tab:CPV} that the short-distance penguin effects are small compared to the tree-amplitude-induced \CP violation. For decay modes with vanishing $a_{dir}^{({\rm tree})}$, the QCD-penguin-induced \CP asymmetry is of order $(1\sim 2)\times 10^{-4}$ or even smaller (see Table~\ref{tab:CPV}).

When the short-distance weak penguin annihilation contribution is estimated in QCDF at the hard-collinear scale $\mu_h\approx$  800 MeV, it is found to be very small as mentioned before. But weak penguin annihilation has the chance to be greatly enhanced. In QCDF, it involves endpoint divergences related to the soft gluon effects. In the diagrammatic approach, it receives long-distance contributions from the color-allowed tree amplitude followed by final-state rescattering as shown in Fig.~\ref{fig:FSI}. Assuming that penguin annihilation is similar to the topological $W$-exchange amplitude [see Eq.~(\ref{eq:PE})], its effect can be studied by comparing $a_{dir}^{({\rm t+pa})}$ with $a_{dir}^{({\rm tree})}$, where the superscript ``pa'' denotes weak penguin annihilation. It is evident from Table~\ref{tab:CPV} that penguin annihilation could play an essential role in the study of \CP violation.

After summing over all possible contributions, we see that the predicted \CP violation denoted by $a_{CP}^{\rm tot}$ (or $a_{CP}^{\rm dir}$) is at most of order $10^{-3}$ in the SM.  For $\Delta a_{CP}^{\rm dir}$, the \CP asymmetry difference in $D^0\to K^+K^-$ and $D^0\to \pi^+\pi^-$, we obtain a value of $-0.13\%$ which is too small compared to the world average of $-(0.645\pm 0.180 )\%$. Since in the SM, $\Delta a_{CP}^{\rm dir}$ stems mainly from weak penguin annihilation, we can vary the amplitude $P\!E$ to see how much enhancement we can gain. Even with the maximal magnitude $|P\!E|\sim T$ and a maximal strong phase relative to $T$, we get $\Delta a_{CP}^{\rm dir}=-0.25\%$. This is more than $2\sigma$ away from the current world average. Hence, if the LHCb result for $\Delta a_{CP}^{\rm dir}$ is confirmed by further data analysis, it will imply new physics in the charm sector.

\section{Conclusions}

In view of the recently measured \CP asymmetry difference $\Delta a_{CP}^{\rm dir}$ between the $D^0 \to K^+ K^-$ and $D^0 \to \pi^+ \pi^-$ modes, we scrutinize the direct \CP violation in the singly Cabibbo-suppressed (SCS) $D \to PP$ and $VP$ decays within the standard model (SM), where $P$ and $V$ refer to pseudoscalar and vector mesons, respectively.  Such an analysis is helpful in diagnosing possible evidence of new physics in the charm sector.

Direct \CP violation in such $D$ decays may arise from the interference between tree-level amplitudes of different topologies.  However, the magnitude is proportional to ${\rm Im}(V_{cd}^*V_{ud} V_{cs}^*V_{us})/|V_{cd}^*V_{ud}|^2$, which is ${\cal O}(10^{-3})$.  Modulated by the size ratio of different flavor amplitudes and the sine of the relative strong phase, we expect that such tree-level \CP asymmetry should be at the order of $10^{-3}$ or even smaller.  The great merit of the topological approach is that the magnitude and the relative strong phase of each individual topological tree amplitude in charm decays can be extracted from the data. Hence, the estimate of $a_{CP}^{\rm dir}$ should be trustworthy and reliable. We find that
$a_{CP}^{\rm dir}(D^0 \to \omega \eta') \sim -1.1 \times 10^{-3}$ is the largest one among all the SCS modes.

In the $D^0 \to K^+ K^-$ and $D^0 \to \pi^+ \pi^-$ decays, the tree-level amplitudes have exactly the same CKM factors and thus do not induce direct \CP asymmetry.  In such cases, one must invoke the penguin amplitudes, including the QCD penguin and weak penguin annihilation amplitudes.  In addition, we have taken into account the SU(3) breaking effects in the color-allowed tree $T$ diagram and the resonance effects in $W$-exchange $E$ diagram, as done in Ref.~\cite{ChengChiang}, in order to explain the observed branching fractions of the two modes.    As given in the $a_{dir}^{(\rm t+p)}$ column of Table~\ref{tab:CPV}, the direct \CP asymmetries of both modes are at a few $\times 10^{-5}$ level.  This is seen to be largely due to the trivial relative strong phase between the QCD penguin amplitude and the tree-level amplitudes.

We next include the contributions of weak penguin annihilation diagrams, $P\!E$, $P\!E_P$ and $P\!E_V$.  The short-distance contributions of such diagrams are typically five orders of magnitude smaller than the color-allowed tree diagram and thus negligible.  However, they may be enhanced by long-distance final-state rescattering through resonances, as in the case of the weak annihilation diagrams $E$ and $A$.  We manage to maximize the \CP-violating effects from the interference with these weak penguin annihilation diagrams by assuming these diagrams are as large as the tree-level $E$, $E_P$, and $E_V$, respectively, and have sizable strong phases relative to $T$.  In the end, we observe that the magnitude of \CP asymmetry in the SCS charm decay modes can reach at most $10^{-3}$.  In particular, the \CP asymmetry difference $\Delta a_{CP}^{\rm did}$ between $D^0 \to K^+ K^-$ and $D^0 \to \pi^+ \pi^-$ is found to be about $-0.13\%$.  Even with the parameter choices of $|PE| \sim |T|$ and a relative strong phase of $90^\circ$, we find $\Delta a_{CP}^{\rm dir}=-0.25\%$, which can be regarded as the upper bound on $\Delta a_{CP}^{\rm dir}$ in the SM, still more than $2\sigma$ away from the world-average experimental result $-(0.645\pm 0.180)\%$.

In conclusion, if $\Delta a_{CP}^{\rm dir}$ continues to be large with more statistics in the future or if the direct \CP asymmetry of any of the discussed modes is significantly larger than $10^{-3}$, it will be clear evidence of physics beyond the SM in the charm sector.

\section*{Acknowledgments}

This research was supported in part by the National Science Council of Taiwan, R.~O.~C. under Grant Nos.~NSC-100-2112-M-001-009-MY3 and NSC-100-2628-M-008-003-MY4 and in part by the NCTS.



\begin{thebibliography}{99}

\bibitem{LHCb}
  R.~Aaij {\it et al.}  [LHCb Collaboration],
  arXiv:1112.0938 [hep-ex].

\bibitem{CDF}
  T.~Aaltonen {\it et al.} [CDF Collaboration],
  arXiv:1111.5023 [hep-ex].

\bibitem{HFAG}
D. Asner {\it et al.} [Heavy Flavor Averaging Group], arXiv:1010.1589 [hep-ex] (2010)
and online update at http://www.slac.stanford.edu/xorg/hfag.

\bibitem{Quigg:1979ic}
  C.~Quigg,
  Z.\ Phys.\ C {\bf 4}, 55 (1980).

\bibitem{Golden:1989qx}
  M.~Golden and B.~Grinstein,
  Phys.\ Lett.\ B {\bf 222}, 501 (1989).

\bibitem{Hinchliffe:1995hz}
  I.~Hinchliffe and T.~A.~Kaeding,
  Phys.\ Rev.\ D {\bf 54}, 914 (1996)
  [hep-ph/9502275].

\bibitem{Ligeti}
  G.~Isidori, J.~F.~Kamenik, Z.~Ligeti and G.~Perez,
  arXiv:1111.4987 [hep-ph].

\bibitem{Kagan}
  J.~Brod, A.~L.~Kagan and J.~Zupan,
  arXiv:1111.5000 [hep-ph].

\bibitem{Zhu}
  K.~Wang and G.~Zhu,
  arXiv:1111.5196 [hep-ph].

\bibitem{Rozanov}
  A.~N.~Rozanov and M.~I.~Vysotsky,
  arXiv:1111.6949 [hep-ph].

\bibitem{Nir}
 Y. Hochberg and Y. Nir, arXiv:1112.5268 [hep-ph].

\bibitem{PU}
 D. Pirktskhalava and P. Uttayarat, arXiv:1112.5451 [hep-ph].

\bibitem{Bigi}
  I.~I.~Bigi and A.~Paul,
  arXiv:1110.2862 [hep-ph].

\bibitem{BigiCDF}
  I.~I.~Bigi, A.~Paul and S.~Recksiegel,
  JHEP\ {\bf 1106}, 089  (2011)  [arXiv:1103.5785 [hep-ph]].


\bibitem{BBNS99}
  M.~Beneke, G.~Buchalla, M.~Neubert and C.~T.~Sachrajda,
  Phys.\ Rev.\ Lett.\  {\bf 83}, 1914 (1999)
  [arXiv:hep-ph/9905312];
  Nucl.\ Phys.\  B {\bf 591}, 313 (2000)
  [arXiv:hep-ph/0006124].


\bibitem{pQCD}
  Y.~Y.~Keum, H.~N.~Li and A.~I.~Sanda,
  Phys.\ Rev.\  D {\bf 63}, 054008 (2001)
  [arXiv:hep-ph/0004173].


\bibitem{SCET}

  C.~W.~Bauer, D.~Pirjol, I.~Z.~Rothstein and I.~W.~Stewart,
  Phys.\ Rev.\  D {\bf 70}, 054015 (2004)
  [arXiv:hep-ph/0401188].


\bibitem{CCBud}
  H.~Y.~Cheng and C.~K.~Chua,
  Phys.\ Rev.\  D {\bf 80}, 114008 (2009);  Phys.\ Rev.\  D {\bf 80}, 074031 (2009).


\bibitem{Cheng:1982}
  H.~Y.~Cheng,
  { Phys.\ Rev.\  D} {\bf 26}, 143 (1982);
  A.~Datta and D.~Kumbhakar,
  { Z.\ Phys.\  C} {\bf 27}, 515 (1985).

\bibitem{Lenz}
  M.~Bobrowski, A.~Lenz, J.~Riedl and J.~Rohrwild,
  { JHEP} {\bf 1003}, 009 (2010)

\bibitem{Falk:y}
  A.~F.~Falk, Y.~Grossman, Z.~Ligeti and A.~A.~Petrov,
  { Phys.\ Rev.\  D} {\bf 65}, 054034 (2002).

\bibitem{CC:mixing}
  H.~Y.~Cheng and C.~W.~Chiang,
  { Phys.\ Rev.\  D} {\bf 81}, 114020 (2010).

\bibitem{Chau}
L. L. Chau Wang, p. 419-431 in AIP Conference Proceedings 72 (1980), Weak Interactions as Probes of
 Unification (edited by G.B. Collins, L.N. Chang and J.R. Ficenec), and p.1218-1232 in Proceedings of
 the 1980 Guangzhou Conference on Theoretical Particle Physics (Science Press, Beijing, China, 1980,
 distributed by Van Nostrand Reinhold company);
  L.~L.~Chau,
  Phys.\ Rept.\  {\bf 95}, 1 (1983).

\bibitem{CC86}
  L.~L.~Chau and H.~Y.~Cheng,
  Phys.\ Rev.\ Lett.\  {\bf 56}, 1655 (1986).

\bibitem{CC87}
  L.~L.~Chau and H.~Y.~Cheng,
  Phys.\ Rev.\  D {\bf 36}, 137 (1987);
  Phys.\ Lett.\  B {\bf 222}, 285 (1989).

\bibitem{ChengOh}
  H.~Y.~Cheng and S.~Oh,
  JHEP\ {\bf 1109}, 024  (2011)  [arXiv:1104.4144 [hep-ph]].

\bibitem{PDG}
 K.~Nakamura {\it et al}. [Particle Data Group], J. Phys. G {\bf 37}, 075021 (2010).

\bibitem{Kass} R. Kass, talk presented at 2009 Europhysics Conference on High Energy Physics, July 16-22, 2009, Krakow, Poland.

\bibitem{ChengChiang}
  H.~Y.~Cheng and C.~W.~Chiang,
  Phys.\ Rev.\ D\ {\bf 81}, 074021  (2010)  [arXiv:1001.0987 [hep-ph]].


\bibitem{RosnerPP08}
  B.~Bhattacharya and J.~L.~Rosner,
  Phys.\ Rev.\  D {\bf 77}, 114020 (2008)
  [arXiv:0803.2385 [hep-ph]].

\bibitem{KLOE}
  F.~Ambrosino {\it et al.},
  JHEP {\bf 0907}, 105 (2009)
  [arXiv:0906.3819 [hep-ph]].

\bibitem{RosnerVP}
  B.~Bhattacharya and J.~L.~Rosner,
  Phys.\ Rev.\  D {\bf 79}, 034016 (2009)
  [arXiv:0812.3167 [hep-ph]].

\bibitem{Zen} P. \.Zenczykowski, Acta Phys. Polon. B {\bf 28}, 1605 (1997).

\bibitem{Weinberg} S. Weinberg, {\it The Quantum Theory of Fields,
Volume I} (Cambridge University Press, New York, 1995), Sec. 3.8.

\bibitem{Chenga1a2} H. Y. Cheng, Eur. Phys. J. C {\bf 26}, 551 (2003).

\bibitem{BN} M. Beneke and M. Neubert, Nucl. Phys. B {\bf 675}, 333
(2003).


\bibitem{BSW} M. Wirbel, B. Stech, and M. Bauer, Z. Phys. C {\bf 29}, 637
(1985); M. Bauer, B. Stech, and M. Wirbel, ibid. C {\bf 34},
103 (1987).

\bibitem{Li}
  H.~-n.~Li and B.~Tseng,
  Phys.\ Rev.\ D {\bf 57}, 443 (1998)  [hep-ph/9706441];
  D.~S.~Du, Y.~Li and C.~D.~Lu,
  Chin.\ Phys.\ Lett.\  {\bf 23}, 2038 (2006)  [hep-ph/0511239].

\bibitem{Du}
  H.~J.~Gong, J.~F.~Sun and D.~S.~Du,
  High Energy Phys.\ Nucl.\ Phys.\  {\bf 26}, 665 (2002)  [hep-ph/0109098].

\bibitem{Wu}
  X.~Y.~Wu, X.~G.~Yin, D.~B.~Chen, Y.~Q.~Guo and Y.~Zeng,
  Mod.\ Phys.\ Lett.\ A {\bf 19}, 1623 (2004);  X.~Y.~Wu, X.~G.~Yin and Y.~Q.~Guo,
  Chin.\ Phys.\  {\bf 13}, 469 (2004);
  X.~Y.~Wu, B.~J.~Zhang, H.~B.~Li, X.~J.~Liu, B.~Liu, J.~W.~Li and Y.~Q.~Guo,
  Phys.\ Lett.\ B {\bf 675}, 196 (2009)  [arXiv:1104.0135 [hep-ph]].

\bibitem{KCYang}
  J.~H.~Lai and K.~C.~Yang,
  Phys.\ Rev.\ D {\bf 72}, 096001 (2005)  [hep-ph/0509092].

\bibitem{Gao}
  D.~N.~Gao,
  Phys.\ Lett.\ B {\bf 645}, 59 (2007)  [hep-ph/0610389].

\bibitem{Grossman}
  Y.~Grossman, A.~L.~Kagan and Y.~Nir,
  Phys.\ Rev.\ D {\bf 75}, 036008 (2007)  [hep-ph/0609178].


\bibitem{Cheng1984}
  L.~L.~Chau and H.~Y.~Cheng,
  Phys.\ Rev.\ Lett.\  {\bf 53}, 1037 (1984).

\bibitem{ChengBud}
  H.~Y.~Cheng and C.~K.~Chua,
  Phys.\ Rev.\ D\ {\bf 80}, 114008  (2009)  [arXiv:0909.5229 [hep-ph]].

\bibitem{Bazavov}
  A.~Bazavov, C.~Bernard, C.~M.~Bouchard, C.~DeTar, M.~Di Pierro, A.~X.~El-Khadra, R.~T.~Evans and E.~D.~Freeland {\it et al.},
  arXiv:1112.3051 [hep-lat].

\bibitem{YLWu}
  Y.~L.~Wu, M.~Zhong and Y.~B.~Zuo,
  Int.\ J.\ Mod.\ Phys.\ A {\bf 21}, 6125 (2006)  [hep-ph/0604007].

\bibitem{Belle}
  E.~Won {\it et al.}  [Belle Collaboration],
  Phys.\ Rev.\ Lett.\  {\bf 107}, 221801 (2011)  [arXiv:1107.0553 [hep-ex]].

\bibitem{Buccella} F. Buccella, M. Lusignoli, G. Miele, A.
Pugliese, and P. Santorelli, Phys. Rev. D {\bf 51}, 3478 (1995).



\end{thebibliography}

\end{document}